\documentclass[aps, twocolumn, secnumarabic, amssymb, nobibnotes, prb, superscriptaddress]{revtex4-2}
\usepackage[utf8]{inputenc}
\usepackage{xcolor}
\usepackage{xcolor}
\usepackage{amsmath}
\usepackage{physics}
\usepackage{graphicx}
\usepackage{wrapfig}
\usepackage{float}
\usepackage{subcaption}
\usepackage{booktabs}
\usepackage{adjustbox}
\usepackage{placeins}
\usepackage{multirow}
\usepackage{array}
\usepackage{setspace}
\usepackage{lineno}
\usepackage{ragged2e} 
\usepackage{caption}
\usepackage[normalem]{ulem}
\graphicspath{{Figs/}}

\def\wberri{\textsc{WannierBerri}}

\newcommand{\yaroslav}[1]{\textcolor{blue}{#1}}
\usepackage{soul}

\usepackage[colorlinks = true, linkcolor = blue, urlcolor  = blue, citecolor = blue, anchorcolor = blue]
{hyperref}
\begin{document}
\title{ Study of the Nonlinear Dependence of Anomalous Hall Conductivity on Magnetization in Weak Itinerant Ferromagnet ZrZn$_2$}
\author{Surasree Sadhukhan}
\email{ssadhukh@gmu.edu}
\affiliation{Department of Physics and Astronomy, George Mason University, Fairfax, VA 22030, USA}
\affiliation{Quantum Science and Engineering Center, George Mason University, Fairfax, VA 22030, USA}
\author{Stepan S. Tsirkin }
\affiliation{Chair of Computational Condensed Matter Physics, Institute of Physics, École Polytechnique Fédérale 
de Lausanne (EPFL), CH-1015 Lausanne, Switzerland}
\author{Yaroslav Zhumagulov}
\affiliation{Chair of Computational Condensed Matter Physics, Institute of Physics, École Polytechnique Fédérale 
de Lausanne (EPFL), CH-1015 Lausanne, Switzerland}
\author{Igor.~I.~Mazin}
\email{imazin2@gmu.edu}
\affiliation{Department of Physics and Astronomy, George Mason University, Fairfax, VA 22030, USA}
\affiliation{Quantum Science and Engineering Center, George Mason University, Fairfax, VA 22030, USA}
\begin{abstract}
As opposed to the ordinary Hall effect, the anomalous Hall effect (AHE) remained unexplained for decades, and, amazingly, some misconceptions have survived even now, in particular, the claim that AHE is linearly related to the net magnetization. Karplus and Luttinger provided a quantum-mechanical explanation of AHE by explicitly including the SOC and the Berry curvature of electronic bands. They did address the question of linearity, but only in the relatively uncommon limit of the exchange coupling smaller than SOC. Now the linear relation in traditional ferromagnets is understood as a domain population effect: both AHE and magnetization are independently proportional to the domain disbalance. In this connection, it is interesting to check to what extent this relation will hold in {\em single-domain} itinerant ferromagnet, the closest case to that analyzed by Karplus and Luttinger? We answer this question by direct calculations, using the Karplus-Luttinger formula, of AHE in a prototypical itinerant ferromagnet, ZrZn$_2$. We show that in the zero-magnetization limit, $M\rightarrow 0$, the linear relation hold, but at rather small moments of $\sim 0.4\ \mu_B$/Zr breaks down completely and even flips the sign.

\end{abstract}
\maketitle
\section*{Introduction}
The ordinary Hall effect has been associated with the transverse voltage appearing due to the Lorentz force acting upon moving electrons in the magnetic field, which was first observed by Edwin Hall in 1879 ~\cite{Hall1879new,Drude1900elektronentheorie, Hurd2012hall}. The movement of electrons in a uniform magnetic field constitutes the main mechanism that produces the classical Hall voltage \cite{Lorentz}. The ordinary Hall effect is a purely classical-charge phenomenon, unrelated to the concept of electron spin, and is not influenced by relativistic SOC ~\cite{Nagaosa2010anomalous, Sinova2015spin} besides small changes in electron dispersion. Later, Edwin Hall found that the transverse-voltage effect is qualitatively different in Fe and other ferromagnets ~\cite{Hall1881}. Initially, it was believed that the internal magnetic field generates an additional Lorentz force, and only much later was it realized that this ``extraordinary'' Hall effect is a purely quantum-mechanical effect and requires that electron have spin (but not necessarily charge).\\
In 1930, an (in)famous formula was first put in writing by E. Pugh ~\cite{Pugh1930},
\begin{equation}
    \sigma_{xy}=R_OH+R_AM, \label{1}
\end{equation}
where $\sigma_{xy}$ is the total Hall conductivity (HC), $\sigma_O=R_OH$ and $\sigma_A=R_AM$ are the ordinary and extraordinary (anomalous) HC, proportional to the applied field and the induced magnetization, respectively. Even though it was realized (much later) that the formula is correct only for multi-domain ferromagnets, where both $M$ and $\sigma_A$ are {\it independently} proportional to the domain population imbalance. Nevertheless, Eq. \ref{1} is even today often used in experimental papers to ``separate'' the ordinary and anomalous contributions. 

The modern description of AHC usually begins with the Karplus--Luttinger (KL) formula ~\cite{Karplus1954hall}. This expression can be viewed as a special case of the general Kubo formula ~\cite{Kubler2012berry, Wang2006ab} (even though it had preceded the latter by several years) use for magnetic system. In non-magnetic centrosymmetric materials, Kramer's degeneracy is preserved, and there is no spin splitting. In ferromagnets, this degeneracy is lifted, and the spin--orbit interaction, which plays the role of an effective internal magnetic field, gives rise to the AHE. Because of this, an external magnetic field or a Lorentz force is not essential for a finite AHC in ferromagnets. This phenomenon is known as the intrinsic anomalous Hall effect, and it arises from the geometry and topology of the electronic band structure. 
The KL formula for AHC can be written as :
\begin{multline}
\sigma^{AHE}_{\alpha\beta}
= \frac{e^{2}}{\hbar}
\sum_{n \ne n'}
\int \frac{d^{3}k}{(2\pi)^{3}}
\,\big[ f(\varepsilon_{n}(\mathbf{k})) - f(\varepsilon_{n'}(\mathbf{k})) \big] \\
\times\,
\operatorname{Im}
\frac{
\langle n,\mathbf{k} \mid v_{\alpha}(\mathbf{k}) \mid n',\mathbf{k} \rangle
\langle n',\mathbf{k} \mid v_{\beta}(\mathbf{k}) \mid n,\mathbf{k} \rangle
}{
\big[ \varepsilon_{n}(\mathbf{k}) - \varepsilon_{n'}(\mathbf{k}) \big]^{2}
}.
\label{Eq1}
\end{multline}
where \( f_{n\mathbf{k}} \) are the Fermi--Dirac distribution functions and
\(\mathbf{v}\) are the velocity operators. 
An equivalent formulation, more often used today, can be given in terms of the Berry curvature ~\cite{Singh2002competition, Xiao2010berry, Sinitsyn2007semiclassical,Jungwirth2002anomalous, Chaudhary2021,Sadhukhan2023,Sadhukhan2025}.

From the mathematical expression above, it might appear that the AHC is directly proportional to the energy difference [$\varepsilon_{n}(\mathbf{k}) - \varepsilon_{n'}(\mathbf{k})$] between the two bands. This has sometimes led to the mistaken idea (inspired by Eq. \ref{1}) that the AHC scales linearly with the exchange splitting. In practice, band crossings formed by majority and minority spin bands near the Fermi level have a much stronger influence on the AHC than the exchange splitting. When the exchange splitting is much smaller than the spin–orbit coupling-induced splitting, such proportionality is restored, but only in this limit. In this context, it is worth noting that a nonzero AHC can emerge even without SOC (the so-called topological, or geometrical, or scalar-choral HE ~\cite{Ohgushi2000spin,Taguchi2001spin,Tatara2002chirality,Taillefumier2006anomalous}), without any net magnetization ~\cite{Smejkal2020}, and even without any exchange splitting ~\cite{SmejkalMazin}. However, these case may be considered by some ``exotic'', ``atypical'' and ``exceptions that confirm the rule''.\\

In this paper, our goal is to show that even in the simplest possible case, a single-domain weak itinerant magnet, ZrZn$_2$, Eq. \ref{1} is only valid in the $M\rightarrow 0$ limit. In fact one of the primary motivations of our investigation is that, after the discovery of altermagnets, it became clear that the AHC can be large even when the net magnetization is zero. Here we therefore test whether invoking such an exotic state is necessary to invalidate this linear formula, or whether it already fails in the simplest possible case. We do it by varying (computationally) the exchange splitting (the itinerant weaK character of ferromagnetism in ZrZn$_2$ makes this protocol physically meaningful) and calculating $\sigma_{xy}$ from Eq. \ref{Eq1}.

\section*{Calculation methodology}

We have investigated the electronic structure and magnetic properties of ZrZn\(_2\) using first-principles density functional theory (DFT). DFT calculations were carried out with the Vienna \textit{ab initio} Simulation Package~(VASP)~\cite{Kresse1996efficient, Kresse1999ultrasoft} and the \yaroslav{GPAW}~\cite{Mortensen2005real,Enkovaara2010electronic,Lehtola2018recent}  codes, both being based on the projector-augmented wave method. 
Exchange–correlation effects were described within the generalized gradient approximation using the PBE functional. In this calculation we have used a plane-wave basis with a 600 eV energy cutoff, and we performed convergence tests with k-point meshes up to \(8 \times 8 \times 8\). 
The structure was optimized using VASP code by relaxing the internal coordinates until all Hellmann–Feynman forces were below \(0.001\) eV/\AA. The electronic self-consistency threshold was set to \(10^{-7}\) eV. 

In this work, our primary aim is to show how the AHC changes as the magnetization varies. To do this, we develop a simple computational approach based on linear interpolation of the Wannier Hamiltonian. 
First, using the the GPAW code we perform self-consistent calculations in the non-magnetic state (zero constrained magnetization), and in the magnetic state, where total energy minimization yields a total magnetic moment of about $1.6\,\mu_\mathrm{B}$ per unit cell (or $0.8\,\mu_\mathrm{B}$ per formula unit). 
Further, we evaluate the electronic bandstructure non-selfconsistently on the irreducible part of the regular 4x4x4 k-grid and construct Wannier functions (WFs) using the  \wberri~\cite{tsirkin2021high} code. SOC in GPAW is treated non-self-consistently\cite{Olsen2016soc}, which in the present calculation is rather an advantage than a drawback, because it allows us to develop a cleaner procedure.
For oth non-magnetic and magnetic states we can construct symmetry-adapted WFs\cite{sakuma_sawf}, separately for spin-up and spin-down channels (for non-magnetic state WFs for the two channels are identical). We use d-states on Zr atoms and s and p states on Zn atoms as initial projections. Because SOC is not included, the WFs are real-valued, and respect all symmetries of the spacegroup 227 (even those tht would be broken by fixing the direction of the magnetization). Further, we extract the magnitude of the SOC Hamiltonian on each atom, project it on the Wannier functions, and add to the Hamiltonian. The ingredients to evaluate the AHC within Wannier interpolation\cite{Wang2006ab} are the matrix elements of the Hamiltonian and position operator:
\begin{eqnarray}
    H_{mn\mathbf{R}} &=& \langle m\mathbf{0}| \hat{H} |n\mathbf{R}\rangle\\
    \mathbf{A}_{mn\mathbf{R}} &=& \langle m\mathbf{0}| \hat{\mathbf{r}} |n\mathbf{R}\rangle
\end{eqnarray}
We obtain them separately for the non-magnetic ($H^0_{mn\mathbf{R}}$, $\mathbf{A}^0_{mn\mathbf{R}}$) and magnetic ($H^1_{mn\mathbf{R}}$, $\mathbf{A}^1_{mn\mathbf{R}}$)  case. Because Wannier functions are similar in the two cases and the gauge is fixed by enforcing the symmetry and phase of the initial projections, it is reasonable to interpolate between the two cases. Thus, we can define 
\begin{eqnarray}
    H_{mn\mathbf{R}} (\alpha) &=& (1-\alpha) H^0_{mn\mathbf{R}} + \alpha H^1_{mn\mathbf{R}}\\
    \mathbf{A}_{mn\mathbf{R}} (\alpha) &=& (1-\alpha) \mathbf{A}^0_{mn\mathbf{R}} + \alpha \mathbf{A}^1_{mn\mathbf{R}}
\end{eqnarray}
Varying $\alpha$ in the rage $[0,1]$, we can evaluate the band-structures, AHC and  the spin magnetization for each value of $\alpha$, following the approach\cite{Wang2006ab} as implemented in\cite{tsirkin2021high}.

The method relies on a Fast Fourier transform of the real-space matrix elements, making use of the crystal symmetries as implemented in \wberri. A dense $k$-mesh $210 \times 210 \times 210$ is used, followed by iterative adaptive refinement to obtain a precise value of the integral of the Berry curvature over the Brillouin zone

\section*{Discussion}


\begin{figure}
    \centering
    \includegraphics[width=0.98\linewidth]{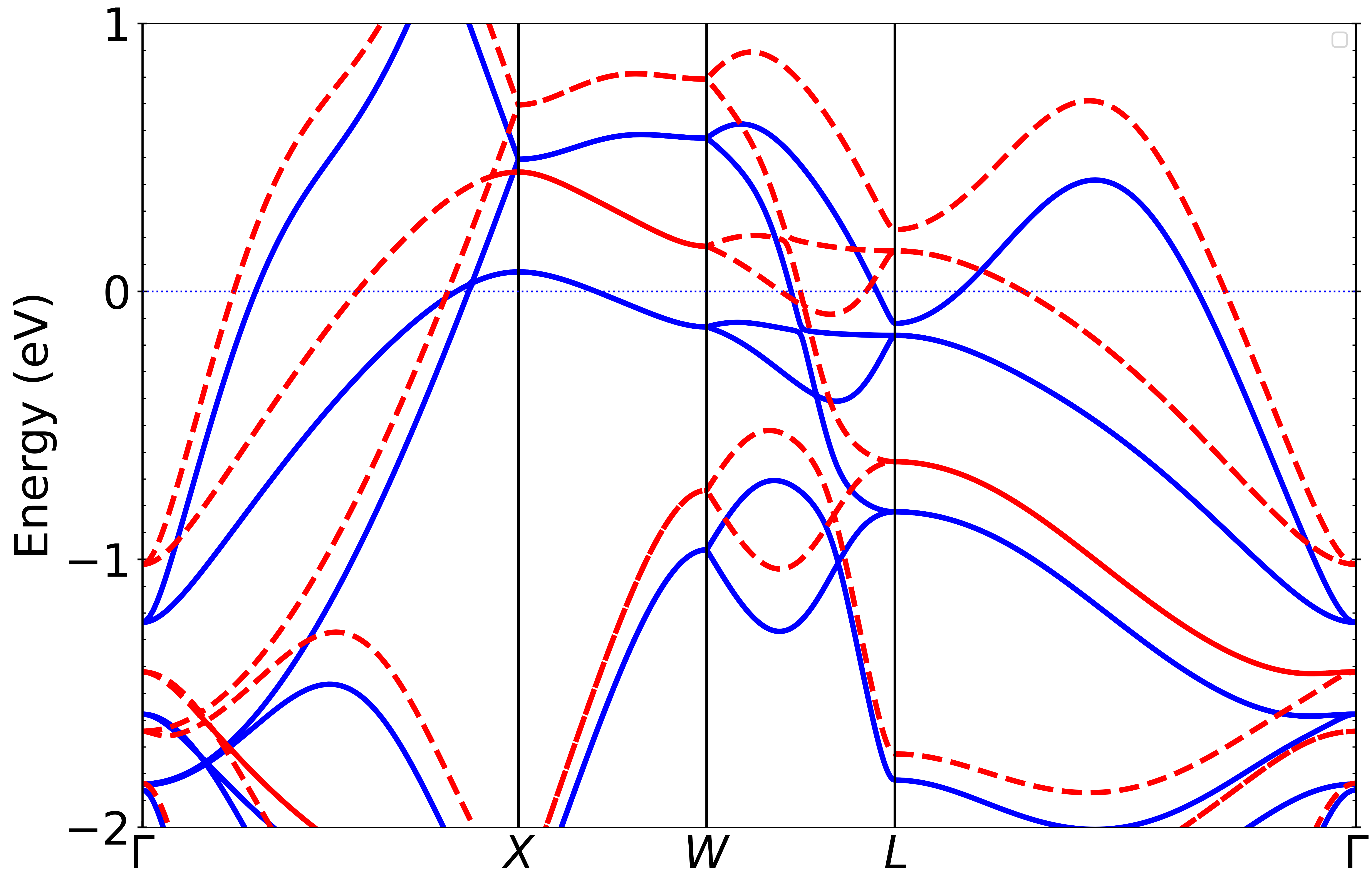}
    \caption{\justifying The figure shows the spin polarized band structure for the majority and the minority spin channel for the ground state FM spin configurations of ZrZn$_2$, respectively. Fermi energy set at 0.0 eV.}
    \label{Fig-1-band}
\end{figure}

\begin{figure*}
\includegraphics[width=1.\textwidth]{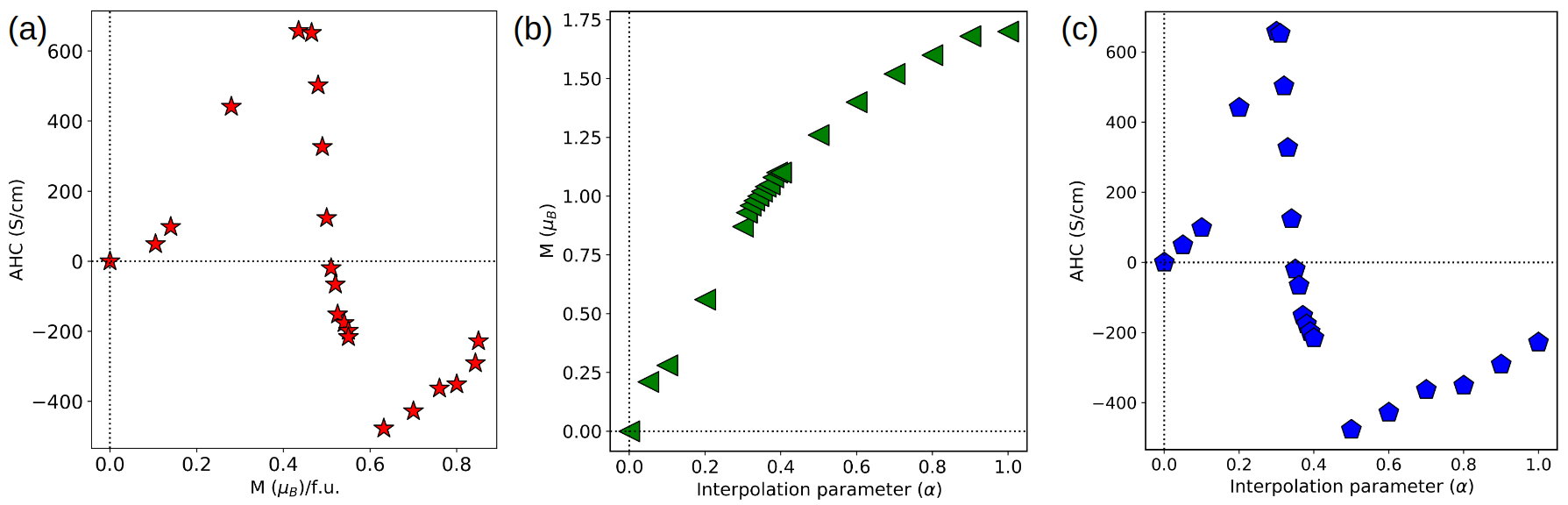}
\captionsetup{width=\textwidth}
\caption{\justifying Figures~(a), (b), and (c) show, respectively, the AHC as a function of the total magnetization, the interpolation parameter ($\alpha$) vs. the total magnetization, the AHC as a function of the total magnetization the interpolation parameter ($\alpha$). The total magnetization is expressed in units of $\mu_{B}$, while the AHC is given in units of S/cm.}
\label{Fig-2-AHC}
\end{figure*}

 ZrZn$_2$ a prototypical weak itinerant ferromagnet, with  magnetism arising from delocalized $4d$ electrons of Zr. The electronic band structure exhibits non-uniform exchange splitting and a rather complex Fermi surface. This makes the low-energy bands rather sensitive to magnetization. ZrZn$_2$ crystallizes in a close-packed
face-centered cubic (FCC) structure with space group \textit{Fd$\bar{3}$m} (No.\ 227) ~\cite{Yelland2005ferromagnetic,Torun2016origin}. The primitive cell contains two Zr atoms and four Zn atoms, corresponding to two formula units of ZrZn$_2$. As typical for itinerant magnets, DFT, being a mean-field theory, overestimates equilibrium magetization even in the local density approximation, to say nothing of gradient-corrected one ~\cite{Mazin2004spin}. Spin fluctuations strongly reduce the static magnetization ~\cite{Moriya2012spin, Moriya1979recent}. These characteristics make ZrZn$_2$ a well-motivated rational choice for us to study the dependence of the AHC on magnetization. \\
The the spin-polarized band structure calculated in the absence of SOC is shown in Fig. \ref{Fig-1-band}. The red-dashed and blue-solid lines represent bands from opposite spin channels. A notable feature is the uneven splitting between the spin-up and spin-down bands across the Brillouin zone, which is particularly evident along the $X-W-L$ path. In this region the spin splitting is relatively large compared to the rest of the band structure. Variations in the spin splitting like these imply a non-uniform magnetism, with a strongly momentum-dependent spin polarization. We have also noticed that some spin-up and spin-down bands approach each other, and overlaps indicate the band mixing and hybridization.\\

In order to study the dependence of AHC on the magnitude of magnetization, we vary the interpolation parameter $\alpha$ from 0 to 1 in steps of 0.1. For every interpolated Hamiltonian, we compute the band structure and evaluate the AHC and total spin magnetization using \wberri. The interpolated band structures  with SOC are shown in supplementary Fig. S1 ~\cite{SM}. We have enlisted the interpolation parameter $\alpha$, the corresponding total magnetization and the AHC  in Supplementary TABLE-1.

In the nonmagnetic case, no SOC-induced splitting is observed because the system respects both inversion and time-reversal symmetries (TRS). For a nonzero $\alpha$ a magnetization is introduced, and due to breaking of TRS the electronic bands start to split [it is not SOC, it is magnetization that lifts degeneracy, even without SOC], and an unequal splitting appears, becoming more pronounced near the $\Gamma$, X, and W points. Moreover, SOC lifts degeneracies where the spin-up and spin-down bands cross in the absence of SOC Around the $\Gamma$ point, particularly near $-1\,\text{eV}$, the magnitude of this splitting increases with increasing $\alpha$. For $\alpha > 0.02$, the bands become more complex due to SOC splitting, and they become denser in the vicinity of the Fermi level. Band rearrangements occur and several bands shift upward, while others move downward with respect to Fermi energy, but the major characteristics remain the same. With increasing $\alpha$, an electron pocket forms, and the resulting band splitting slightly modifies the band topology, making more complex Fermi-surface nesting due to band hybridization, and changes the 
number of band crossings at the Fermi energy. All these factors can influence transport properties specially AHC as we can see in this work. Far above the Fermi level, beyond $1\,\text{eV}$, the overall band dispersion broadens.\\
The calculated AHC for different interpolation parameters is shown as a function of magnetization in Fig.~\ref{Fig-2-AHC}(a). 

We have also illustrated the Berry curvature distribution across the electronic band structure, where positive and negative contributions are represented by red and blue colors, respectively, as shown in Supplementary Fig.~S3. In the nonmagnetic case, the Berry curvature vanishes identically, and consequently the AHC is zero. With an elevated value of the interpolation parameter $\alpha$, the interpolated system breaks TRS, thus giving rise to a nonzero Berry curvature in momentum space. The difference between the positive and negative  Berry curvature gives rise to a non zero AHC value. The sign and value of AHC completely depend on the net Berry curvature present in the Brillouin zone.\\
 
As observed from Fig.~\ref{Fig-2-AHC}(a), it is clear that the anomalous Hall effect is not a linear function of magnetization for ZrZn$_2$ and our result agrees with the modern theoretical understanding of the AHC. The recent understanding focuses on the role of the Berry curvature and band topology or geometry decides the nature of AHC. It does not maintain a linear dependence or direct connection with exchange splitting or magnetization which was a popular misconception in condensed matter community. We also note an essential feature in Fig.~\ref{Fig-2-AHC}(a): the AHC first increases and then decreases as the magnetization increases and AHC approaches a negative value near a total moment of $1.10\,\mu_{\mathrm{B}}$ which corresponds to an interpolation parameter $\alpha$=0.4. It is important to mention that the total magnetization does not follow a linear variation with the interpolation parameter $\alpha$ as shown in  Fig.~\ref{Fig-2-AHC}(b). \\
In Fig.~\ref{Fig-2-AHC}(c), we present the AHC as a function of the interpolation parameter~$\alpha$. To obtain more profound insight into the emergence of the negative-to-positive AHC, we computed the AHC for a denser set of interpolation parameters between $\alpha = 0.30$ and $0.40$, with excellent sampling in the region $\alpha = 0.31$--$0.39$. These refined calculations suggest that a topological phase transition may take place within the narrow interval between $\alpha = 0.34$ and $\alpha = 0.35$. This transition seems to occur only through a change in the geometric arrangement of the electronic bands, without breaking the symmetry of the system or introducing any order parameter, and hence is conceptually quite different from a conventional Landau-type phase transition. In order to provide further insights into the underlying mechanism, we have studied the behavior of the electronic band structure in this range. In Supplementary Fig.~S3, band structures for $\alpha = 0.34$, $0.35$, and $0.36$ are presented. It is clear that the gap closes at $\alpha = 0.34$ and reopens at $\alpha = 0.36$, indicating toward an unconventional topological phase transition. The point at which the gap closes is highlighted with a red circle in Supplementary Fig.~S3, and this identifies the key feature of the band structure responsible for the topological character change.

\section*{Summary and Conclusion}
In this study, we calculated AHC in weak itinerant ferromagnet ZrZn$_2$ in order to demonstrate that the AHC does not depend on magnetization linearly in a single domain soft magnet, when the amplitude of the magnetic moment varies. We find that in the small magnetization limit, $M\rightarrow 0$, the linearity is recovered, consistent with the Karplus-Luttinger conclusion ~\cite{Karplus1954hall}. However, at $M\sim 0.15\ \mu_B$/Zr the slope changes, and at  $M\sim 0.4--0.6\ \mu_B$/Zr changes sign. These dramatic deviations from the usually assumed linearity are due to changes in the electronic structure, in particularly, in the latter case, to a Lifshits transition. Thus, we demonstrated that even in one of the simplest possible cases the standard prescription of Eq. \ref{1} does not work.\\

\section*{Acknowledgments}
This work was supported by the Army Research Office under Cooperative Agreement Number W911NF-22-2-0173. We also acknowledge support from QSEC and computational resources from the Hopper HPC cluster at GMU. 

\bibliography{bib}

@article{sakuma_sawf,
  title = {Symmetry-adapted Wannier functions in the maximal localization procedure},
  author = {Sakuma, R.},
  journal = {Phys. Rev. B},
  volume = {87},
  issue = {23},
  pages = {235109},
  numpages = {8},
  year = {2013},
  month = {Jun},
  publisher = {American Physical Society},
  doi = {10.1103/PhysRevB.87.235109},
  url = {https://link.aps.org/doi/10.1103/PhysRevB.87.235109}
}

@article{Hall1879new,
  title={On a new action of the magnet on electric currents},
  author={Hall, Edwin H.},
  journal={American Journal of Mathematics},
  volume={2},
  pages={287--292},
  year={1879}
}

@article{Drude1900elektronentheorie,
  title={Zur Elektronentheorie der Metalle},
  author={Drude, Paul},
  journal={Annalen der Physik},
  volume={306},
  pages={566--613},
  year={1900}
}

@book{Hurd2012hall,
  title={The Hall Effect in Metals and Alloys},
  author={Hurd, Charles},
  publisher={Springer},
  year={2012}
}

@article{Lorentz,
  title={La th{\'e}orie {\'e}lectromagn{\'e}tique de Maxwell et son application aux corps mouvants},
  author={Lorentz, Hendrik A.},
  journal={Archives N{\'e}erlandaises des Sciences Exactes et Naturelles},
  volume={25},
  pages={363--552},
  year={1892}
}

@article{Nagaosa2010anomalous,
  title={Anomalous Hall effect},
  author={Nagaosa, Naoto and Sinova, Jairo and Onoda, Shigeki and MacDonald, Allan H. and Ong, N. P.},
  journal={Reviews of Modern Physics},
  volume={82},
  pages={1539--1592},
  year={2010}
}

@article{Sinova2015spin,
  title={Spin Hall effects},
  author={Sinova, Jairo and Valenzuela, Sergio O. and Wunderlich, J{\"o}rg and Back, Christian H. and Jungwirth, Tomas},
  journal={Reviews of Modern Physics},
  volume={87},
  pages={1213--1260},
  year={2015}
}

@article{Hall1881,
  title={On the rotational coefficient in nickel and cobalt},
  author={Hall, Edwin H.},
  journal={Philosophical Magazine},
  volume={12},
  pages={157--172},
  year={1881}
}

@article{Pugh1930,
  title={Hall effect and the magnetic properties of some ferromagnetic materials},
  author={Pugh, Emerson M.},
  journal={Physical Review},
  volume={36},
  pages={1503--1511},
  year={1930}
}

@article{Karplus1954hall,
  title={Hall effect in ferromagnetics},
  author={Karplus, Robert and Luttinger, Joaquin M.},
  journal={Physical Review},
  volume={95},
  pages={1154},
  year={1954}
}

@article{Kubler2012berry,
  title={Berry curvature and the anomalous Hall effect in Heusler compounds},
  author={K{\"u}bler, J{\"u}rgen and Felser, Claudia},
  journal={Physical Review B},
  volume={85},
  pages={012405},
  year={2012}
}

@article{Wang2006ab,
  title = {Ab initio calculation of the anomalous Hall conductivity by Wannier interpolation},
  author = {Wang, Xinjie and Yates, Jonathan R. and Souza, Ivo and Vanderbilt, David},
  journal = {Phys. Rev. B},
  volume = {74},
  issue = {19},
  pages = {195118},
  numpages = {15},
  year = {2006},
  month = {Nov},
  publisher = {American Physical Society},
  doi = {10.1103/PhysRevB.74.195118},
  url = {https://link.aps.org/doi/10.1103/PhysRevB.74.195118}
}

@article{Singh2002competition,
  title={Competition of spin fluctuations and phonons in superconductivity of {ZrZn$_2$}},
  author={Singh, David J. and Mazin, I. I.},
  journal={Physical Review Letters},
  volume={88},
  pages={187004},
  year={2002}
}

@article{Xiao2010berry,
  title={Berry phase effects on electronic properties},
  author={Xiao, Di and Chang, Ming-Che and Niu, Qian},
  journal={Reviews of Modern Physics},
  volume={82},
  pages={1959--2007},
  year={2010}
}

@article{Sinitsyn2007semiclassical,
  title={Semiclassical theories of the anomalous Hall effect},
  author={Sinitsyn, N. A.},
  journal={Journal of Physics: Condensed Matter},
  volume={20},
  pages={023201},
  year={2007}
}

@article{Jungwirth2002anomalous,
  title={Anomalous Hall effect in ferromagnetic semiconductors},
  author={Jungwirth, Tomas and Niu, Qian and MacDonald, Allan H.},
  journal={Physical Review Letters},
  volume={88},
  pages={207208},
  year={2002}
}

@article{Chaudhary2021,
  title   = {Role of chemical disorder in tuning the Weyl points in vanadium-doped {Co$_2$TiSn}},
  author  = {Chaudhary, P. and Dubey, K. K. and Shukla, G. K. and Singh, S. and Sadhukhan, S. and Kanungo, S. and Jena, A. K. and Lee, S.-C. and Bhattacharjee, S. and Min{\'a}r, J.},
  journal = {Physical Review Materials},
  volume  = {5},
  number  = {12},
  pages   = {124201},
  year    = {2021},
  doi     = {10.1103/PhysRevMaterials.5.124201}
}

@article{Sadhukhan2023,
  title   = {Atomistic designing of 2D quantum materials heterostructures {CdF/CrI$_3$} for Berry curvature driven tunable intrinsic anomalous Hall state},
  author  = {Sadhukhan, S. and Kanungo, S.},
  journal = {Journal of Physics: Condensed Matter},
  volume  = {35},
  number  = {45},
  pages   = {455601},
  year    = {2023},
  doi     = {10.1088/1361-648X/acee7d}
}

@article{Sadhukhan2025,
  title={Nonsymmorphic symmetry enforced Weyl nodal metal in chalcopyrite {MnGeAs$_2$}},
  author={Sadhukhan, S. and Kanungo, S.},
  journal={Scientific Reports},
  volume={15},
  pages={34739},
  year={2025}
}

@article{Ohgushi2000spin,
  title={Spin anisotropy and quantum Hall effect in the kagom{\'e} lattice},
  author={Ohgushi, K. and Murakami, S. and Nagaosa, N.},
  journal={Physical Review B},
  volume={62},
  pages={R6065},
  year={2000}
}

@article{Taguchi2001spin,
  title={Spin chirality, Berry phase, and anomalous Hall effect in a frustrated ferromagnet},
  author={Taguchi, Y. and Oohara, Y. and Yoshizawa, H. and Nagaosa, N. and Tokura, Y.},
  journal={Science},
  volume={291},
  pages={2573--2576},
  year={2001}
}

@article{Tatara2002chirality,
  title={Chirality-driven anomalous Hall effect in the weak coupling regime},
  author={Tatara, G. and Kawamura, H.},
  journal={Journal of the Physical Society of Japan},
  volume={71},
  pages={2613--2616},
  year={2002}
}

@article{Taillefumier2006anomalous,
  title={Anomalous Hall effect due to spin chirality in the kagom{\'e} lattice},
  author={Taillefumier, M. and Canals, B. and Lacroix, C. and Dugaev, V. K. and Bruno, P.},
  journal={Physical Review B},
  volume={74},
  pages={085105},
  year={2006}
}

@article{Smejkal2020,
  title={Crystal time-reversal symmetry breaking and spontaneous Hall effect in collinear antiferromagnets},
  author={Smejkal, Libor and Gonzalez-Hernandez, Ricardo and Jungwirth, Tomas and Sinova, Jairo},
  journal={Science Advances},
  volume={6},
  pages={eaaz8809},
  year={2020}
}

@unpublished{SmejkalMazin,
  title={Is it possible to sort out ordinary, anomalous and topological Hall effects?},
  author={Smejkal, Libor and Mazin, I. I.},
  note={Unpublished}
}

@article{Kresse1996efficient,
  title={Efficient iterative schemes for ab initio total-energy calculations using a plane-wave basis set},
  author={Kresse, Georg and Furthm{\"u}ller, J{\"u}rgen},
  journal={Physical Review B},
  volume={54},
  pages={11169},
  year={1996}
}

@article{Kresse1999ultrasoft,
  title={From ultrasoft pseudopotentials to the projector augmented-wave method},
  author={Kresse, Georg and Joubert, Daniel},
  journal={Physical Review B},
  volume={59},
  pages={1758},
  year={1999}
}

@article{Mortensen2005real,
  title={Real-space grid implementation of the projector augmented wave method},
  author={Mortensen, Jens J{\o}rgen and Hansen, Lars B. and Jacobsen, Karsten W.},
  journal={Physical Review B},
  volume={71},
  pages={035109},
  year={2005}
}

@article{Enkovaara2010electronic,
  title={Electronic structure calculations with GPAW},
  author={Enkovaara, J. and others},
  journal={Journal of Physics: Condensed Matter},
  volume={22},
  pages={253202},
  year={2010}
}

@article{Lehtola2018recent,
  title={Recent developments in Libxc},
  author={Lehtola, S. and others},
  journal={SoftwareX},
  volume={7},
  pages={1--5},
  year={2018}
}

@article{Tsirkin2021high,
  title={High performance Wannier interpolation of Berry curvature and related quantities with WannierBerri code},
  author={Tsirkin, Stepan S.},
  journal={npj Computational Materials},
  volume={7},
  pages={33},
  year={2021}
}

@article{Yelland2005ferromagnetic,
  title={Ferromagnetic properties of {ZrZn$_2$}},
  author={Yelland, E. A. and others},
  journal={Physical Review B},
  volume={72},
  pages={184436},
  year={2005}
}

@article{Torun2016origin,
  title={Origin of weak magnetism in compounds with cubic Laves structure},
  author={Torun, E. and Janner, A. and de Groot, R. A.},
  journal={Journal of Physics: Condensed Matter},
  volume={28},
  pages={065501},
  year={2016}
}

@article{Mazin2004spin,
  title={Spin fluctuations and the magnetic phase diagram of ZrZn$_2$},
  author={Mazin, I. I. and Singh, David J.},
  journal={Physical Review B},
  volume={69},
  pages={020402},
  year={2004}
}

@book{Moriya2012spin,
  title={Spin Fluctuations in Itinerant Electron Magnetism},
  author={Moriya, T{\^o}ru},
  publisher={Springer},
  volume={56},
  year={2012}
}

@article{Moriya1979recent,
  title={Recent progress in the theory of itinerant electron magnetism},
  author={Moriya, T{\^o}ru},
  journal={Journal of Magnetism and Magnetic Materials},
  volume={14},
  pages={1--46},
  year={1979}
}

@article{SM, 
title={Supplementary Material: The supplementary material contains the interpolated bands with SOC and Berry curvature corresponding to different interpolation parameters $\alpha$.}
}

@article{Olsen2016soc,
  title = {Designing in-plane heterostructures of quantum spin Hall insulators from first principles: $1{\mathrm{T}}^{\ensuremath{'}}\ensuremath{-}{\mathrm{MoS}}_{2}$ with adsorbates},
  author = {Olsen, Thomas},
  journal = {Phys. Rev. B},
  volume = {94},
  issue = {23},
  pages = {235106},
  numpages = {9},
  year = {2016},
  month = {Dec},
  publisher = {American Physical Society},
  doi = {10.1103/PhysRevB.94.235106},
  url = {https://link.aps.org/doi/10.1103/PhysRevB.94.235106}
}

\end{document}